\documentclass{article}
\usepackage[english]{babel}
\usepackage[utf8]{inputenc}
\usepackage{johd}

\title{\textbf{When Revisiting is Wrong! \\ Rebuttal: Revisiting Neural Program Smoothing
for Fuzzing}}
\author{Dongdong She$^{\ast}$, Kexin Pei$^{\dagger}$, Junfeng Yang$^{\star}$, Baishakhi Ray$^{\star}$, Suman Jana$^\star$  \\
  \small $^\ast$Hong Kong University of Science and Technology \\
  \small $^\dagger$The University of Chicago\\ 
  \small $^\star$Columbia University \\
}

\date{}
\usepackage{comment}
\usepackage{listings}
\usepackage{xcolor}
\usepackage{booktabs} 
\definecolor{codegreen}{rgb}{0,0.6,0}
\definecolor{codegray}{rgb}{0.5,0.5,0.5}
\definecolor{codepurple}{rgb}{0.58,0,0.82}
\definecolor{backcolour}{rgb}{0.95,0.95,0.92}
\definecolor{highlight}{RGB}{255,0,0} 

\lstdefinestyle{mystyle}{
    backgroundcolor=\color{backcolour},   
    commentstyle=\color{codegreen},
    keywordstyle=\color{magenta},
    numberstyle=\tiny\color{codegray},
    stringstyle=\color{codepurple},
    basicstyle=\ttfamily\footnotesize,
    breakatwhitespace=false,         
    breaklines=true,                 
    captionpos=b,                    
    keepspaces=true,                 
    numbers=left,                    
    numbersep=5pt,                  
    showspaces=false,                
    showstringspaces=false,
    showtabs=false,                  
    tabsize=2
}

\begin{document}
\maketitle


\begin{abstract} 

MLFuzz, a work accepted at ACM FSE 2023, revisits the performance of a machine learning-based fuzzer, NEUZZ. We demonstrate that its main conclusion is entirely wrong due to several fatal bugs in the implementation and wrong evaluation setups, including an initialization bug in persistent mode, a program crash, an error in training dataset collection, and a mistake in fuzzing result collection.
Additionally, MLFuzz uses noisy training datasets without sufficient data cleaning and preprocessing, which contributes to a drastic performance drop in NEUZZ. We address these issues and provide a corrected implementation and evaluation setup, showing that NEUZZ consistently performs well over AFL on the FuzzBench dataset. Finally, we reflect on the evaluation methods used in MLFuzz and offer practical advice on fair and scientific fuzzing evaluations.

\end{abstract}
\section{Introduction}
NEUZZ is a machine learning-based fuzzer that leverages a neural network model to boost fuzzing performance~\cite{she2018neuzz}. MLFuzz~\cite{Maria2023} is a recent revisiting work on NEUZZ that claims the performance of prior work NEUZZ does not hold. It also gives an in-depth analysis of the limitations of machine learning-based fuzzers. In this paper, we show that MLFuzz's major claim is completely wrong due to severe fatal bugs in its implementation and experimental setup errors, specifically as follows:
\begin{itemize}
  \item An initialization bug $\Longrightarrow$ Failure setup of persistent mode fuzzing.
  \item A program crash $\Longrightarrow$ Unexpected early termination of NEUZZ.
  \item An error in training dataset collection $\Longrightarrow$ A poorly-trained neural network model.
  \item An error in NEUZZ's result collection $\Longrightarrow$ Incomplete code coverage report of NEUZZ.
\end{itemize}
These implementation bugs and experimental errors lead to unexpected performance downgrades in NEUZZ, resulting in inaccurate analyses and conclusions. 
We refute its in-depth analysis of the machine learning-based fuzzer and show that they are wrong and misleading to the fuzzing community.

However, MLFuzz was accepted at ACM FSE 2023, one of the top-tier software engineering conferences. In this short paper, we first explain the technical details of the persistent mode initialization bug in MLFuzz's implementation. We then analyze the Python program crash in MLFuzz's neural network module caused by careless API misuse. Subsequently, we highlight the mistakes MLFuzz made when collecting the training dataset for NEUZZ. We also point out an error in MLFuzz's experimental setup that leads to incomplete test cases and fuzzing coverage collection. 
These implementation bugs, along with experimental setup errors, led to invalid evaluation results of NEUZZ in the MLFuzz paper and invited further wrong in-depth analysis of the machine learning-based fuzzer. We refute its analysis with a case study on ML-based fuzzer speed. In the end,
we fixed the bug in MLFuzz's implementation and properly configured NEUZZ on the Google FuzzBench dataset. Our preliminary evaluation demonstrates that NEUZZ's performance gain over vanilla AFL on the FuzzBench programs is consistent with what was reported in its original paper. We summarize the lessons learned from MLFuzz's mistakes and propose several tips for correctly and scientifically evaluating and revisiting fuzzing works.

\section{Initialization Bug in Persistent Mode}
\label{sec:init_bug}
MLFuzz introduces an initialization bug that causes a failure in setting up the persistent mode of NEUZZ. Its implementation of NEUZZ then degenerates into a non-persistent mode. Furthermore, all the evaluation of NEUZZ in MLFuzz turns into an \underline{\emph{invalid}} apple-to-banana comparison (i.e., comparison of high-speed persistent mode fuzzing vs. low-speed non-persistent mode fuzzing), whose result is dominated by the significant gap between persist mode and non-persistent mode rather than the fuzzing algorithm difference. 

\subsection{Persistent Mode Fuzzing}
Persistent mode is a high-speed mode of fuzzing that can offer 10X or 20X execution speedup~\cite{persistentAFL} compared with non-persistent mode. It encapsulates the entire tested project into a function and repetitively invokes that function within a persistent loop. Each iteration of the function call is provided with a testcase, representing a single execution instance of the tested project. While the non-persistent fuzzing mode has to fork a new process for each testcase, incurring a large runtime overhead. Moreover, the non-persistent fuzzing mode transfers the test case generated by the fuzzer through the slow disk I/O, significantly slower than the high-speed shared memory used in persistent mode. To sum up, persistent mode fuzzing offers two levels of performance speedup: 1) test case execution: through a function call vs. through a new process fork; 2) test case transmission: high-speed memory vs. low-speed disk I/O. With these benefits, persistent mode fuzzing enjoys a higher fuzzing throughput and better performance than conventional fuzzing mode. 


\begin{figure}[H]
\centering
\includegraphics[width=0.45\textwidth]{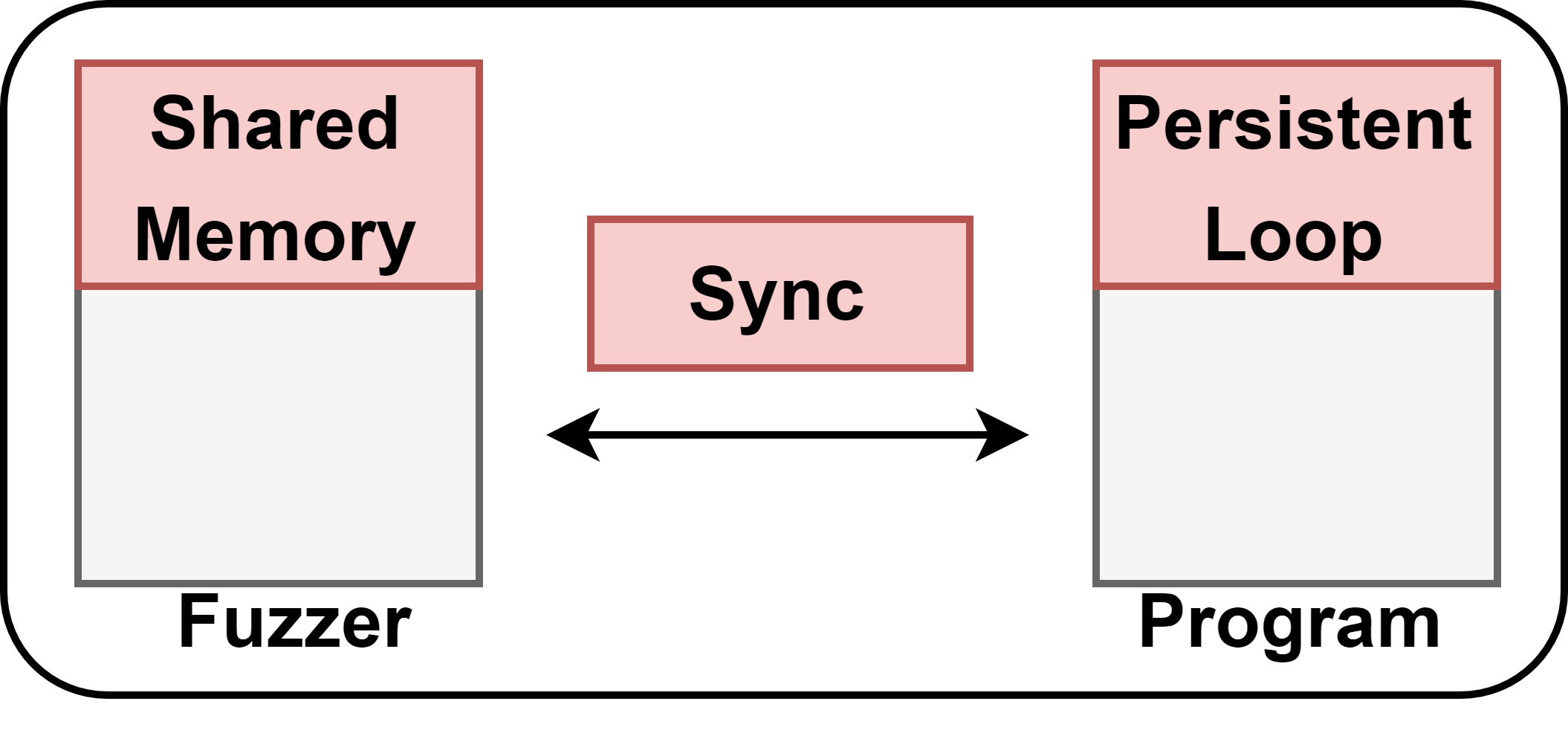}
\caption{\small\textbf{Overview of Persistent Mode Design. We highlight the three major components in \textcolor{red}{red}.}}
\label{fig:overview}
\end{figure}
\subsection{Persistent Mode Design}
We briefly overview the persistent mode's design to help readers understand MLFuzz's initialization bug. Figure~\ref{fig:overview} shows the high-level architecture of the persistent mode fuzzing. It consists of three main components highlighted in \textcolor{red}{red} rectangle. The fuzzer on the left side implements a shared memory I/O that enables the high-speed test case transmission throughout a chunk of shared memory between the fuzzer and the tested program. On the right side, the program includes a persistent loop where each iteration invokes a function call encapsulating the entire tested program. In the middle, the persistent mode fuzzing synchronizes the state of the fuzzer and the tested program with customized environment variables. The synchronization functions as a switch. It enables the persistent mode fuzzing only when the fuzzer and program are set up properly. Otherwise, it downgrades into non-persistent mode fuzzing.


\lstset{style=mystyle}

\subsection{MLFuzz's Initialization Bug}
The vanilla NEUZZ only supports the non-persistent fuzzing mode and is \underline{\emph{incompatible}} with the persistent mode. MLFuzz runs NEUZZ in persistent mode by introducing its own patch. However, the patch contains a fatal bug that causes a failure in the persistent mode setup. As a result, their implementation of NEUZZ degenerates into low-speed, non-persistent fuzzing mode. 

We show the bug in a simplified code snippet in Listing ~\ref{lst:bug}. Line 3 defines a state variable \texttt{is\_persistent} shared between fuzzer and program, indicating whether the persistent mode is set or not. By default, fuzzers like AFL, AFL++ assume persistent mode is off. Lines 5-12 denote fuzzer end code implementation, and lines 14-17 denote program end code implementation. The fuzzer first tries initializing the synchronization state variable \texttt{is\_persistent} by checking whether the tested program supports persistent mode. Then, it sets up the shared memory I/O accordingly. Similarly, the program sets up the persistent loop based on the value of the synchronization state variable. The initialization bug occurs in line 8 when MLFuzz forgot to set the state variable value \texttt{is\_persistent}. As a result, the state variable turns into a constant false. Both \texttt{shared\_memory\_IO()} in fuzzer end and \texttt{persistent\_loop()} in program end are not properly invoked as expected. The setup of persistent mode fails in MLFuzz's evaluation. This initiation bug makes all the evaluation results of NEUZZ on the FuzzBench dataset become invalid and meaningless.
\begin{lstlisting}[language=C, caption={
\small\textbf{The initialization bug in MLFuzz's persistent mode impelmentation. MLFuzz forgets to set the synchronization state variable is\_persistent at line 8. The state variable turns into a constant false, leading to a failure in the persistent mode setup. }}, label={lst:bug}]
/* mlfuzz/fuzzers/neuzz_experiments/neuzz.c_fixed_seed.patch */
/* Define a synchronization state variable */
bool is_persistent; 

Fuzzer-end source code:
  (*@\textbf{\textcolor{highlight}{/* Initialization BUG: MLFuzz forgets to set the value of state variable at line 8. */}}@*)
  (*@\textbf{\textcolor{highlight}{/* As a result, is\_persistent is  mistakenly set as a constant false value. */}}@*)
  (*@\textbf{\textcolor{highlight}{// is\_persistent = check\_binary();}}@*)
  ...
  /* Setup shared-memory I/O if the state variable is true */
  if (is_persistent) 
    shared_memory_IO();

Program-end source code:
  /* setup persistent loop if the state variable is true */
  if (is_persistent)
    persistent_loop();
\end{lstlisting}
\subsection{Wrong In-depth Analysis on the Speed of ML-based Based Fuzzer}
We refute a factual error in MLFuzz's in-depth analysis on the speed of ML-based fuzzer. MLFuzz claims that ML-based fuzzers are orders of magnitude slower than other fuzzers regarding execution throughput. They also justified that claim with experimental evidence in their paper section 5.7. However, that evaluation is completely invalid due to the initialization bug in their code, as shown in section ~\ref{sec:init_bug}. As a matter of fact, they were comparing NEUZZ in the non-persistent mode against other fuzzers in the persistent mode. The orders of magnitude execution throughput is caused by the speed optimization of persistent mode over non-persistent mode. With a fatal bug in the code, their evaluation turns into a typical apple-to-banana comparison. Note that ML-based fuzzers are mutation algorithms with minimal overhead and an orthogonal component to the persistent mode. Ultimately, the bogus evaluation leads to a completely wrong and misleading conclusion that ML-based fuzzers are orders of magnitude slower than other fuzzers. Our preliminary evaluation in Table~\ref{table:speed} shows that ML-based fuzzers with the correct persistent mode setup have around the same fuzzing throughput as other fuzzers.
\section{Crash Bug in Python Program.}
MLFuzz's patch introduces another bug that leads to a crash of the Python module of NEUZZ during the fuzzing campaign. This ignored crash further causes an early termination of NEUZZ during their 24-hour evaluation. Even though NEUZZ did not complete the 24-hour evaluation, MLFuzz mistakenly reported the early termination number as the final result for NEUZZ. Such a careless and irresponsible evaluation setup results in invalid and bogus results.   

As shown in code listing~\ref{lst:crash}, \texttt{gen\_adv2()} and \texttt{gen\_adv3()} are two functions computing the gradient for a given input using the \emph{same} Keras API~\footnote{https://keras.io/}. In line 7, the function \texttt{gen\_adv2()} calls the correct API \texttt{tf.keras.\colorbox{highlight}{backend}.function()} to instantiate a Keras function. However, in line 15, the same API usage is mistaken as \texttt{tf.keras.\colorbox{highlight}{gradient}.function()}. We mark the difference in red to highlight the wrong API usage. Specifically, MLFuzz mistakes the API \texttt{backend} as \texttt{gradient}. Such invalid API usage directly invites a crash in the neural network module implemented in Python. It then results in an unexpected early termination of NEUZZ during the fuzzing campaign. We are surprised that MLFuzz's patch lacks basic syntax checking when launching scientific evaluations on prior works.

\lstset{style=mystyle}
\begin{lstlisting}[language=Python, caption={\small\textbf{Crash Bug in NN Module Introduced by MLFuzz. Wrong API usage at line 15 leads to an unexpected program crash.}}, label={lst:crash}]
# mlfuzz/fuzzers/neuzz_experiments/nn.py_tf2.patch
# Compute gradient for a given input
def gen_adv2(...):
  ...
  grads = tf.keras.backend.gradients(...)
  (*@\textbf{\textcolor{highlight}{\# Correct API usage at line 7}}@*)
  iterate = tf.keras.(*@\colorbox{highlight}{backend}@*).function(...)
  ...

# Compute gradient for a given input without sign
def gen_adv3(...):
  ...
  grads = tf.keras.backend.gradients(...)
  (*@\textbf{\textcolor{highlight}{\# Crash BUG: wrong API usage at line 15 leading to a crash}}@*)
  iterate = tf.keras.(*@\colorbox{highlight}{gradient}@*).function(...)  
  ...
\end{lstlisting}


\section{Error in Dataset Collection}
In addition to the implementation bugs, we found several fatal errors in the machine-learning model setup of NEUZZ from the MLFuzz artifact. In this section, we show that MLFuzz carelessly collected an extremely noisy training dataset without any data preprocessing and cleaning, leading to an unexpectedly poor-performance neural network model. Specifically, its dataset is drastically variant, unbalanced, and contains insufficient samples. 
The training data includes testcases from a few bytes to millions of bytes. The original NEUZZ assumes all test cases in the training data have around the same file length, and each test case only has a few bytes different from each other. Only with such a clean and strictly aligned dataset can the small neural network model in NEUZZ learn useful patterns. As a common machine learning principle states \underline{\emph{garbage in, garbage out}}. An unexpectedly noisy training dataset causes an invalid and useless neural network model in NEUZZ. Therefore, its further in-depth analysis of the neural network model is misleading due to the fact that the model is not even properly trained in the first place. 

The proper training dataset collection for NEUZZ requires running AFL deterministic mode for an hour with a \underline{\emph{single}} seed. The benefit of such a design is that the testcases generated by the one-hour fuzzing are all variant testcases with a small magnitude of perturbations, centering around the same seed file. Although the total number of testcases collected during the one-hour fuzzing is relatively small (i.e., range from 1k-10k), they only have a slight difference from each other. Hence, the small neural network model can converge on this small, simple training dataset.   

However, MLFuzz ignores such guidelines. They carelessly use a seed corpus consisting of \underline{\emph{multiple}} seeds to feed into the one-hour fuzzing to collect the training data samples. Recent works have shown that the choice of seed corpus significantly impacts the resulting test case generated during a fuzzing campaign~\cite{Seed2021Adrian, wolff2022explainable}. After running AFL default mode with multiple seeds for an hour, the generated testcases (i.e., collected training data samples) have drastic differences from each other since they are derived from different seed files. The complex and messy dataset consists of insufficient samples. Such a noisy dataset leads to poor neural network model convergence. 

\section{Error in Fuzzing Result Collection}
MLFuzz makes a mistake when collecting NEUZZ's fuzzing result corpus. The original implementation of NEUZZ saves fuzzing result testcases into \underline{\emph{two separate}} directories based on the file length of test cases. Nevertheless, MLFuzz only collects the test cases in one directory while ignoring the other. As a result, the edge coverage of NEUZZ reported in MLFuzz is incomplete. 
The core mutation scheme of NEUZZ is a fixed-length mutation, i.e., all test cases generated have the exact same file length as its parent test case. All the fixed-length test cases generated are saved into a directory named \texttt{NEUZZ\_out}. In addition, NEUZZ employs a variant length mutation scheme based on an insertion or deletion mutation operator. These test cases with variant file lengths are saved into a separate directory named \texttt{vari\_seed}. MLFuzz carelessly ignored the testcases in \texttt{vari\_seed} and reported a incomplete number in their evaluation. 
\section{Preliminary Result}
We fixed the implementation bugs and experimental setup errors described above in MLFuzz. We then evaluate the performance of NEUZZ against vanilla AFL on the Google FuzzBench dataset. We ensure both fuzzers are \underline{\emph{properly}} set up in the persistent mode and run for 24 hours. We opensource our artifact on Github \footnote{\url {https://github.com/Dongdongshe/mlfuzz/tree/main/fuzzers/neuzz_experiments}}. The preliminary result is shown in Table~\ref{table:eval}. The results demonstrate that NEUZZ significantly outperforms vanilla AFL on the four Google FuzzBench programs, consistent with the result reported in the NEUZZ paper. We further measure the fuzzing throughput (i.e., the number of executions per second) of NEUZZ and vanilla AFL. Table~\ref{table:speed} shows that the speed of NEUZZ is around the same as vanilla AFL. The wrong and misleading result in MLFuzz~\cite{Maria2023} is caused by the persistent mode initialization bug, crash bug, and other errors in its experimental setup. 
\begin{table}[!ht]
\caption{\small\textbf{Mean edge coverage of NEUZZ against AFL on four Google FuzzBench programs over 5 runs for 24 hours.}}
    \centering
    \label{table:eval}
    \begin{tabular}{lrr}
    \toprule
        \textbf {Program} & \textbf{NEUZZ} & \textbf{AFL} \\ 
        \midrule
 bloaty    &  5,626  &     5,092  \\
 curl      &  9,086  &     8,878  \\
 freetype  &  8,267  &     6,157  \\
 libxml    &  8,433  &     6,619  \\
        \bottomrule
    \end{tabular}
\end{table}
\begin{table}[!ht]
\caption{\small\textbf{Mean fuzzing throughput of NEUZZ against AFL on four Google FuzzBench programs over 5 runs for 24 hours.}}
    \centering
    \label{table:speed}
    \begin{tabular}{lrr}
    \toprule
        \textbf {Program} & \textbf{NEUZZ} & \textbf{AFL} \\ 
        \midrule
 bloaty    &  470  &     445  \\
 curl      &  1,832  &     1,808  \\
 freetype  &  2,167  &     2,201  \\
 libxml    &  3,363  &     3,308  \\
        \bottomrule
    \end{tabular}
\end{table}
\section{Lessons Learnt from MLFuzz's Mistakes}
Critiques are crucial to scientific discovery and advancement. Yet, wrong critique can only lead to misleading and confusing conclusions. We introspect the erroneous experimental setups and wrong conclusions in MLFuzz~\cite{Maria2023}, then summarize the following valuable lessons for fuzzing researchers and practitioners when working on revisiting and evaluation projects.

\subsection{Tips for Correct Implementation} 
Fuzzing research needs to launch large-scale fuzzing campaigns, i.e., compare the performance of multiple open-source fuzzers under a specific experimental setting. Often, researchers might have to modify or patch an open-source fuzzer project. For instance, MLFuzz patches NEUZZ to support persistent mode in order to evaluate it under a new dataset -- Google FuzzBench. However, its careless and irresponsible implementation contains several fatal bugs, leading to completely bogus evaluation results and conclusions. Learning from their mistakes, we propose the following two suggestions: 
\begin{itemize}
  \item \textbf{Careful and Responsible Debugging.} Researchers need to examine new codes introduced into a prior work carefully. An unnoticed bug or design flaw could easily cause unexpected performance decrease in prior work. Therefore, we suggest researchers responsibly debug code patches added to prior works and ensure their implementation is correct. Meanwhile, the newly added code patch should not hurt the performance of prior work.
  \item \textbf{Seeking Help from Authors.} If the researchers have difficulty patching a prior work. It is a professional move to contact the original authors for help. We believe the authors of MLFuzz lack a basic understanding of AFL persistent mode and NEUZZ design after careful examination of their implementation. However, as the authors of NEUZZ, we never received any email from MLFuzz's authors seeking help about the persistent mode.  
\end{itemize}

\subsection{Tips for Correct Experimental Setup}
Conducting a fair and scientific fuzzing evaluation is hard. A subtle error in the experimental setup can cripple a fuzzer's performance. We summarize a list of tips that contribute to correct experimental setups.   
\begin{itemize}
    \item \textbf{Diverse and Representative Benchmarks.} There exist many fuzzer benchmark datasets, such as Google FuzzBench~\cite{fuzzbench}, Magma~\cite{magma}, and UniBench~\cite{unibench}. Prior work shows that a specified fuzzer can have inconsistent results on different benchmark programs (i.e., there is no silver button fuzzer that can win the championship on all the benchmark programs)~\cite{unifuzz}. A diverse and representative benchmark is crucial for a fair and scientific fuzzing evaluation. However, MLFuzz evaluates NUEZZ on a single dataset--the FuzzBench with which NEUZZ is originally \underline{\emph{incompatible}}. Although the set of programs and artifacts evaluated in the NEUZZ paper have been open-sourced on GitHub, MLFuzz ignores them without any justification. Note that the FuzzBench dataset was released in 2020, two years after the publication of NEUZZ. MLFuzz further overclaims that NEUZZ performance does not hold without specifying a particular benchmark. Their evaluation cannot sufficiently justify their claim. Reflecting on MLFuzz's mistake, we suggest fuzzing researchers responsibly and carefully choosing a diverse and representative dataset in fuzzer evaluation. 
    \item \textbf{Uniform Code Coverage Metric.} Code coverage implementation varies on fuzzers. AFL implements an XOR-based hash edge coverage that is prone to hash collision~\cite{afl}. LLVM designs a collision-free code coverage called LLVM CoverageSanitizer that by default prunes some redundant nodes from the control flow graph~\cite{covsanit}. K-Scheduler uses a non-prune feature LLVM CoverageSanitizer~\cite{She2022EffectiveSS}. 
    AFL++ implements its inline-optimized LLVM CoverageSanitizer~\cite{aflpp}. A fair and scientific evaluation must convert these different code metrics into a uniform one through test case replay. 
    \item \textbf{Complete Testcase Collection.} Fuzzers can save testcases into different directories depending on specific design choices. AFL saves slow testcases into \texttt{hang} directory and crash testcases into \texttt{crash} directory. NEUZZ would save all variant-length mutations into a separate directory \texttt{vari\_seeds}. However, MLFuzz forgot to collect the testcase from \texttt{vari\_seeds} in their evaluation of NEUZZ, leading to incomplete results. A proper code coverage replay should include testcases from all output directories.
    \item \textbf{Uniform Fuzzing Mode.} Persistent mode can significantly boost fuzzing performance using persistent loop and shared memory I/O optimization. The fatal initialization bug in MLFuzz's implementation leaves NEUZZ in a non-persistent mode while all other fuzzers are set up in persistent mode. Researchers should double-check that all evaluated fuzzers are running in the same mode. 
    \item \textbf{Open-source Fuzzing Seed Corpus.} Fuzzing is essentially a search. Without a given starting starting point, it's almost impossible to reproduce the search process. We encourage fuzzer researchers to open-source the seed corpus and the fuzzer in the reproduction artifact. 
\end{itemize}
~\
\section{Conclusion}
MLFuzz is a revisiting of machine learning-based fuzzers, accepted at ACM FSE 2023. It mainly claims that NEUZZ's performance over AFL does not hold. We show such a conclusion is completely wrong due to several fatal bugs in its implementation bugs and experimental setup. Meanwhile, based on invalid evaluation observation, MLFuzz's in-depth analysis of ML-based fuzzers is also misleading. We fixed its bug and experimental setup and demonstrated that the performance of NEUZZ over vanilla AFL on the FuzzBench dataset still holds. We reflect on the mistakes of MLFuzz and summarize lessons on scientific and fair fuzzing evaluation and revisiting.

\bibliographystyle{acm}
\bibliography{bib}

\end{document}